\documentclass{PoS}

\usepackage{default}
\usepackage{slashed}

\title{Minimally doubled fermions at one-loop level}

\ShortTitle{Minimally doubled fermions at one-loop level}

\author{Stefano Capitani\\
        Institute for Nuclear Physics, Johannes Gutenberg University, Mainz\\
        E-mail: \email{capitan@kph.uni-mainz.de}}

\author{\speaker{Johannes Weber}
	\thanks{
 We thank Mike Creutz for useful discussions. This work was supported by
 Deutsche Forschungsgemeinschaft (SFB443), Gesellschaft f\"ur
 Schwerionenforschung GSI and the Research Center ``Elementary Forces \& Mathemetical Foundations''.}\\
       Institute for Nuclear Physics, Johannes Gutenberg University, Mainz\\
       E-mail: \email{weberj@kph.uni-mainz.de}}

\author{Hartmut Wittig\\
        Institute for Nuclear Physics, Johannes Gutenberg University, Mainz\\
        E-mail: \email{wittig@kph.uni-mainz.de}}

\abstract{Single fermionic degrees of freedom together with standard chiral
 symmetry at finite lattice spacing, correct continuum limit and local
 interactions only are precluded by the Nielsen-Ninomiya no-go theorem.
 The class of minimally doubled fermion actions exhibits exactly two
 chiral modes. Recent interest in these actions has been sparked by the
 investigation of fermionic actions defined on ``hyperdiamond'' lattices. 

 Due to the necessity of breaking hypercubic symmetry explicitly,
 radiative corrections generate operator mixings with relevant and
 marginal operators that should vanish in continuum QCD. These cannot
 be avoided and must be taken into account in particular by a peculiar
 wave-function renormalisation and additive momentum renormalisation.

 Renormalisation properties at one-loop level of the self-energy, local
 bilinears and conserved vector and axial-vector currents are presented
 for Bori\c ci-Creutz and Karsten-Wilczek actions. Distinct differences
 and similarities between both actions are elucidated.}

\FullConference{The XXVII International Symposium on Lattice Field Theory - LAT2009\\
		 July 26-31 2009\\
		 Peking University, Beijing, China}

\begin{document}

\section{Introduction}

 For many years since the early days of lattice QCD, chiral symmetry was
 regarded as incompatible with lattice regularisation. The Nielsen-Ninomiya
 no-go theorem forbids the existence of a single chiral mode, which has a
 correct continuum limit as well as local interactions only.
 

 Minimally doubled fermions represent a class of actions, which exactly
 satisfy the minimal requirements of the no-go theorem. A prominent
 representative of the minimally doubled fermion class is the
 Bori\c ci-Creutz Dirac operator \cite{1,2}:
 \begin{equation}
  D_{BC}(k) = i \sum\limits_{\mu} \tilde k_{\mu} \gamma_{\mu}
 -\frac{i}{2} a \sum\limits_{\mu} \hat k_{\mu}^{2} \gamma_{\mu}^{\,\prime} +m_{0}.
 \label{eq:DBC}
 \end{equation}
 The trigonometric functions of the lattice momenta are defined as usual
 and the second set of gamma matrices is defined by a relation, which
 breaks hypercubic symmetry:
 \begin{eqnarray}
  \tilde k_{\mu} \equiv \frac{1}{a} \sin (a k_{\mu}),
 & \hat k_{\mu} \equiv \frac{2}{a} \sin (a \frac{k_{\mu}}{2}),
 \label{eq:lattice ks}\\
  \gamma_{\mu}^{\,\prime} \equiv \Gamma \gamma_{\mu} \Gamma = \Gamma -\gamma_{\mu},
 & 2 \Gamma \equiv \sum\limits_{\mu} \gamma_{\mu} = \sum\limits_{\mu} \gamma_{\mu}^{\,\prime}.
 \label{eq:gammas}
 \end{eqnarray}
 This Dirac operator posesses two zeros. Their nature is made transparent, when the
 Bori\c ci-Creutz term is cast into another form:
 \begin{equation}
  -\frac{i}{2} a \sum\limits_{\mu} \hat k_{\mu}^{2} \gamma_{\mu}^{\,\prime} =
  i \frac{1}{a} \sum\limits_{\mu} \cos (a k_{\mu}) \gamma_{\mu}^{\,\prime} - 2i \frac{1}{a} \Gamma.
 \label{eq:BCterm}
 \end{equation}
 The cosine functions are reduced to unity at $k = (0,0,0,0)$. Application of
 (\ref{eq:gammas}) clearly shows that both parts of the Bori\c ci-Creutz term
 on the right hand side of (\ref{eq:BCterm}) compensate at this point. On the
 other hand, the first sum in (\ref{eq:DBC}) evaluated at
 \mbox{$k = (\frac{\pi}{2a},\frac{\pi}{2a},\frac{\pi}{2a},\frac{\pi}{2a})$}
 compensates the second half of the Bori\c ci-Creutz term, while the cosine
 functions in its first half vanish. Other zeros do not exist. Both are
 situated on the hypercubic main diagonal, which is the symmetry breaking
 axis. A combined symmetry transformation in all components
 \begin{equation}
  k_{\mu} \to \frac{\pi}{2a}-k_{\mu},\quad
  \left(\begin{array}{c}
   \gamma_{\mu} \\                                        
   \gamma_{\mu}^{\,\prime}
  \end{array}\right)
  \to
  \left(\begin{array}{c}
   \gamma_{\mu}^{\,\prime} \\                                        
   \gamma_{\mu}
  \end{array}\right)
 \label{eq:BCsymmetry}
 \end{equation}
 does not change the Bori\c ci-Creutz Dirac operator, but changes the sign of the chirality
 matrix: $\gamma_{5}^{\,\prime}=\Gamma \gamma_{5} \Gamma = -\gamma_{5}$. It corresponds to an
 exchange of the poles which have opposite chirality.


\section{Perturbation theory for Bori\c ci-Creutz fermions}

\subsection{Propagators and vertices}

 Our recent study \cite{3} of Bori\c ci-Creutz fermions proved the occurence of operator
 mixings due to one-loop effects. Effects of this sort had been conjectured \cite{4}
 previously. Here, we revisit the properties of Bori\c ci-Creutz fermions and compare them later
 on to Karsten-Wilczek fermions.

 The propagator is obtained from the inversion of the Dirac operator \cite{3,5}:
 \begin{equation}
  S_{BC} = \frac{ -i \sum\limits_{\mu} \tilde k_{\mu} \gamma_{\mu}
   +\frac{i}{2} a \sum\limits_{\mu} \hat k_{\mu}^{2} \gamma_{\mu}^{\,\prime} +m_{0}}
  {\sum\limits_{\mu} \hat k_{\mu}^{2}+a \sum\limits_{\mu} \tilde k_{\mu}\big(\hat k_{\mu}^{2}
   - \frac{1}{2}\sum\limits_{\nu} \hat k_{\nu}^{2} \big) + m_{0}^{2}}.
 \label{eq:SBC}
 \end{equation}
 The violation of hypercubic symmetry is obvious, as the denominator cannot be cast into
 a form with definite behaviour under reversal of any direction.

 The weak coupling expansion of the gauge field
 $U_{\mu}(x)=e^{ i a g_{0} A_{\mu}(x+\frac{a}{2}e_{\mu})}$ is performed in the usual
 manner \cite{3,5}. Quark vertices with one or two gluons are denoted by $V^{1}$ and $V^{2}$:
 \begin{eqnarray}
  V_{\mu}^{1}(p_{1},p_{2}) &=& -i g_{0} \Big( \gamma_{\mu} \cos \frac{(p_{1}+p_{2})_{\mu}}{2} 
  -\gamma_{\mu}^{\,\prime} \sin \frac{(p_{1}+p_{2})_{\mu}}{2} \Big),
 \label{eq:v1BC} \\
  V_{\mu}^{2}(p_{1},p_{2}) &=&
  i \frac{a}{2} g_{0}^{2} \Big( \gamma_{\mu} \sin \frac{(p_{1}+p_{2})_{\mu}}{2} 
  +\gamma_{\mu}^{\,\prime} \cos \frac{(p_{1}+p_{2})_{\mu}}{2} \Big).
 \label{eq:v2BC}
 \end{eqnarray}
 These vertices can be derived from Wilson fermion vertices by the replacement
 \mbox{$r f_{\mu} \to -i\gamma_{\mu}^{\,\prime} f_{\mu}$}. Due to the subtle
 difference that the Dirac structure of $-i\gamma_{\mu}^{\,\prime} f_{\mu}$ is
 different for each $\mu$, even the evaluation of simple diagrams is very
 complex and generates vast numbers of terms.

\subsection{Self-energy}

 Two diagrams\footnote{For a list of the diagrams see Fig. 1 in \cite{3}.} add up
 to the self-energy at one-loop level. Due to the Dirac structure of the n-point
 functions, computation of the lattice integrals requires evaluation of every
 possible combination of indices of Dirac matrices and momenta.

 The tadpole diagram's contribution,
 \begin{equation}
  g_{0}^{2}C_{F} \frac{Z_{0}}{2} \Big(1 - \frac{1}{4}(1-\alpha) \Big)
  \Big(i \slashed{p} +2i\frac{1}{a} \Gamma \Big),
 \label{eq:BCtadpole}
 \end{equation}
 with $Z_{0} = 24.466100 / (16\pi^{2})$, contains a power-divergent part. The
 possibility that this power-divergence might be canceled by the sunset diagram
 is not realised \cite{4} . Nevertheless, gauge invariance requires at least a cancellation
 of the part proportional to $(1-\alpha)$.

 Evaluation of the sunset diagram yields
 \begin{eqnarray}
  & \,\,\,\,\,\,\,\,\,\, i \slashed{p} \cdot\frac{g_{0}^{2} C_{F}}{16\pi^{2}}
   & \Big( \log a^{2}p^{2}-5.42642 +(1-\alpha) \big( -\log a^{2}p^{2}+7.850272 \big) \Big) \\
  & + m_{0} \cdot\frac{g_{0}^{2} C_{F}}{16\pi^{2}}
   & \Big( 4\log a^{2}p^{2}-29.48729 +(1-\alpha) \big( -\log a^{2}p^{2}+5.792010 \big) \Big) \\
  & \!\!+ i \Gamma \Pi \! \cdot \! \frac{g_{0}^{2} C_{F}}{16\pi^{2}} & \!\!\!\cdot\,\,\, 1.52766
  \label{eq:BChsbp}\\
  & \! + i\frac{1}{a} \Gamma \! \cdot \! \frac{g_{0}^{2} C_{F}}{16\pi^{2}}
   & \Big( 5.07558 +6.11653 (1-\alpha) \Big).
  \label{eq:BCpd}
 \label{eq:BCsunset}
 \end{eqnarray}
 Herein, we used the definition $\Pi\equiv \sum\limits_{\mu}p_{\mu}$. Thus,
 the structure in (\ref{eq:BChsbp}) is proportional to the momentum projection
 on the hypercubic symmetry-breaking axis. It can be cast into a more transparent
 form by using anticommutation relations for Dirac matrices including $\Gamma$:
 \begin{equation}
  \Gamma \Pi = \frac{1}{2}\{\Gamma,\{\Gamma,\slashed{p}\}\}=\slashed{p}+\slashed{p}^{\,\prime},
 \label{eq:BCtransparence}
 \end{equation}
 with $\slashed{p}^{\,\prime} \equiv \sum\limits_{\mu} p_{\mu} \gamma_{\mu}^{\,\prime}$.

 Furthermore, (\ref{eq:BCpd}) cancels the power-divergent part of (\ref{eq:BCtadpole})
 that is forbidden by gauge invariance. The further power-divergences, however, not only fail to cancel, but amplify each
 other.

 The full one-loop expression for the self energy is
 \begin{equation}
  \Sigma(p,m_{0}) = i \slashed{p} \Sigma_{1}(p) + m_{0} \Sigma_{2}(p)
  + c_{1}(g_{0}^{2}) i (\slashed{p}+\slashed{p}^{\,\prime})  + c_{2}(g_{0}^{2}) i \frac{1}{a} \Gamma,
  \label{eq:BCselfenergy}
 \end{equation}
 with
 \begin{eqnarray}
  & \Sigma_{1}(p) =
   & 1+\frac{g_{0}^{2}C_{F}}{16 \pi^{2}}\Big( \log a^{2}p^{2} + 6.80663
     +(1-\alpha) \big( -\log a^{2}p^{2}+4.792010 \big)\Big), \\
  & \Sigma_{2}(p) =
   & 1+\frac{g_{0}^{2}C_{F}}{16 \pi^{2}}\Big( 4\log a^{2}p^{2} -29.48729 
     +(1-\alpha) \big( -\log a^{2}p^{2}+5.792010 \big)\Big), \\
  & c_{1}(g_{0}^{2}) =
   & 1.52766 \cdot\frac{g_{0}^{2}C_{F}}{16 \pi^{2}}, \\
  & c_{2}(g_{0}^{2}) =
   & 29.54170 \cdot\frac{Cg_{0}^{2}C_{F}}{16 \pi^{2}}. \label{eq:BCc2}
 \end{eqnarray}
 Obviously, (\ref{eq:BCtransparence}) must enter into the wave-function
 renormalisation in a non-trivial way. Therefore,
 $Z_{\psi} = Z_{\psi}(\Sigma_{1}(p), c_{1}(g_{0}^{2}))$.

\subsection{Local bilinears}

 The renormalisation of local bilinears is a straigthforward procedure. As
 chiral symmetry demands, scalar and pseudoscalar densities as well as 
 local vector and axial-vector currents have the same renormalisation factors.
 Without taking the wave-function renormalisation into account, the proper
 renormalisation factors read
 \begin{eqnarray}
  & \Lambda_{S}(p) =
   & \frac{C_{f} g_{0}^{2}}{16 \pi^{2}}\Big( -4\log a^{2}p^{2} +29.48729
     +(1-\alpha) \big( \log a^{2}p^{2}-5.792010 \big)\Big), \\
  & \Lambda_{V}(p) =
   & \frac{C_{f} g_{0}^{2}}{16 \pi^{2}} \Big( -\log a^{2}p^{2} +9.54612 
     +(1-\alpha) \big( \log a^{2}p^{2}-4.792010 \big)\Big), \\
  & \Lambda_{T}(p) =
   & \frac{C_{f} g_{0}^{2}}{16 \pi^{2}} \Big( 2.16548 
     +(1-\alpha) \big( \log a^{2}p^{2}-3.792010 \big)\Big).
 \end{eqnarray}
 The local vector and axial-vector currents suffer from an additional
 operator mixing besides the wave-function renormalisation:
 \begin{equation}
  \overline{\psi} \,\gamma_{\mu} \psi \to
  \overline{\psi}^{R} \gamma_{\mu} \,\big(1+ Z_{\psi} + \Lambda_{V}(p) \big) \psi^{R}
  + c^{vtx}(g_{0}^{2})\overline{\psi}^{R} \,\Gamma \psi^{R},
 \label{eq:BClvc}
 \end{equation}
 with $c^{vtx}(g_{0}^{2})=-0.10037\cdot\frac{g_{0}^{2}C_{F}}{16 \pi^{2}}$.
 Since each coordinate axis has a non-vanishing projection on the hypercubic
 main diagonal, the symmetry breaking operator mixes with each of the four
 components. The nature of this mixing can be visualised by applying
 (\ref{eq:gammas}): $\Gamma = \gamma_{\mu}+\gamma_{\mu}^{\,\prime}$.

\section{One-loop properties}

\subsection{Momentum renormalisation}

 Due to the fact that it is proportional to a Dirac gamma matrix, the
 power-divergence in the self energy is unlike its counterpart in the
 Wilson case. The mass is protected from additive renormalisation as
 chiral symmetry is unbroken. Instead, the four-momentum is subject to
 renormalisation:

 \begin{equation}
  \check p_{\mu} = p_{\mu} -\frac{c_{2}(g_{0}^{2})}{2a} \quad\rightarrow\quad
  \slashed{p} = \slashed{\check p} + \frac{c_{2}(g_{0}^{2})}{a} \Gamma.
 \label{eq:BCmomentumrenormalisation}
 \end{equation}
 The conjecture that quark velocities had to renormalised \cite{2} is thus verified.
 Since neither pole lies at $(0,0,0,0)$ any more, the definition of a quark
 rest frame becomes non-trivial. Besides that issue, the relevant quantity
 for periodic boundaries is $c_{2}(g_{0}^{2})$ modulo $2\pi$.

\subsection{Conserved currents}

 After the Bori\c ci-Creutz action is cast into coordinate space, 
 \begin{eqnarray}
  S_{BC} = 
   & a^{4} \sum\limits_{x}
     \Big(\sum\limits_{\mu} \frac{1}{2a} \Big(\overline{\psi}(x)
     (\gamma_{\mu}+i \gamma_{\mu}')U_{\mu}(x)\psi(x+a\,e_{\mu})
     -\overline{\psi}(x+a\,e_{\mu})(\gamma_{\mu}-i \gamma_{\mu}')
     U_{\mu}^{\dagger}(x)\psi(x) \Big) \nonumber\\
   & \ \ +\overline{\psi}(x) \big( m_{0}-2i \frac{1}{a} \Gamma \big)
     \psi(x) \Big),
 \label{eq:cspaceSBC}
 \end{eqnarray}
 conserved vector and axial-vector currents can be derived by application of the
 Ward identities \cite{6}. The transformations
 \begin{equation}
  \left(\begin{array}{c}
   \psi(x) \\
   \overline{\psi}(x)
  \end{array}\right)
  \to 
  \left(\begin{array}{c}
  (1 + i \alpha_{V}) \psi(x) \\
  \overline{\psi}(x) (1 - i \alpha_{V})
  \end{array}\right), \quad
  \left(\begin{array}{c}
   \psi(x) \\
   \overline{\psi}(x)
  \end{array}\right)
  \to 
  \left(\begin{array}{c}
  (1 + i \alpha_{A} \gamma_{5}) \psi(x) \\
  \overline{\psi}(x) (1 + i \alpha_{A} \gamma_{5} )
  \end{array}\right)
 \label{eq:chiraltransformations}
 \end{equation}
 yield conserved point-split currents
 \footnote{Obviously, the axial-vector current is conserved only in the chiral limit.}:
 \begin{eqnarray}
  V_{\mu}^{c}(x) = &
   \frac{1}{2} \Big(\overline{\psi}(x)(\gamma_{\mu}+i \gamma_{\mu}^{\,\prime})
   U_{\mu}(x)\psi(x+a\, e_{\mu})
   +\overline{\psi}(x+a\, e_{\mu})(\gamma_{\mu}-i \gamma_{\mu}^{\,\prime})
   U_{\mu}^{\dagger}(x)\psi(x) \Big), \\
  A_{\mu}^{c}(x) = &
   \frac{1}{2} \Big(\overline{\psi}(x)(\gamma_{\mu}+i \gamma_{\mu}^{\,\prime})
   \gamma_{5} U_{\mu}(x)\psi(x+a\, e_{\mu})
   +\overline{\psi}(x+a\, e_{\mu})(\gamma_{\mu}-i \gamma_{\mu}^{\,\prime})
   \gamma_{5} U_{\mu}^{\dagger}(x)\psi(x) \Big).
 \label{eq:BCcurrents}
 \end{eqnarray}

 Four diagrams\footnote{The diagrams are listed in Fig. 1 in \cite{3}.} contribute to their renormalisation: vertex diagram,
 operator tadpole and two sails. In the case of the vector current, the
 proper current renormalisation amounts to
 \begin{eqnarray}
  & \Lambda_{V^{c}}(p) = &
   \frac{g_{0}^{2}C_{F}}{16 \pi^{2}}\Big( -\log a^{2}p^{2} - 6.80664 + (1-\alpha) \big( \log a^{2}p^{2}-4.792010 \big)\Big), \nonumber\\
  & c_{V^{c}}(g_{0}^{2}) = &
   -1.52766 \cdot\frac{g_{0}^{2}C_{F}}{16 \pi^{2}}.
 \label{eq:BCcvcrenormalistion}
 \end{eqnarray}
 The full expression for the renormalisation of the conserved vector current is
 \begin{equation}
  Z_{V^{c}} \overline{\psi}\gamma_{\mu}\psi =
  \Big( (1 + Z_{\psi}) \overline{\psi}^{R}\gamma_{\mu}\psi^{R}
   + \overline{\psi}^{R}( \Lambda_{V^{c}}(p) \gamma_{\mu} + c_{V^{c}}(g_{0}^{2})
    \Gamma )\psi^{R}\Big).
 \end{equation}
 It is not straightforward to proof that $Z_{V^{c}}$ is unity.
 $\Gamma = \gamma_{\mu}+\gamma_{\mu}^{\,\prime}$, which is obtained from
 (\ref{eq:gammas}), reveals the same structure as (\ref{eq:BCtransparence}).
 Furthermore, the renormalisation factors inside the wave-function renormalisation
 ($\Sigma_{1}(p)$, $c_{1}(g_{0}^{2})$) have signs exactly opposite to those in the
 proper current renormalisation ($\Lambda_{V^{c}}(p)$, $c_{V^{c}}(g_{0}^{2})$). It
 seems reasonable to assume that these structures cancel, even though an algebraic
 proof does not exist yet.

\section{Karsten-Wilczek fermions}

 A full study of the one-loop properties of Karsten-Wilczek fermions \cite{7} is in
 preparation \cite{8}. The Karsten-Wilczek Dirac operator,
 \begin{equation}
  S_{KW}(k) = i \sum\limits_{\mu} \tilde k_{\mu} \gamma_{\mu}
 +\frac{i}{2} \lambda a \sum\limits_{\mu} \hat k_{\mu}^{2} \gamma_{4}(1-\delta_{\mu4}) +m_{0},
 \end{equation}
 has two zeros located at $k = (0,0,0,0)$ and $k = (0,0,0,\frac{\pi}{a})$. It contains
 the Wilczek parameter $\lambda$, which is constrained to $|\lambda| \geq \frac{1}{2}$.
 The cancellation of the additional zeros could not be achieved otherwise. A second
 set of gamma matrices can be defined:
 \begin{equation}
  \gamma_{\mu}^{\,\prime} \equiv
  \gamma_{4} \gamma_{\mu} \gamma_{4} = (2\delta_{\mu4}-1) \gamma_{\mu}.
 \label{eq:KWgammaprime}
 \end{equation}

 Due to the Kronecker symbol, the Karsten-Wilczek term includes only the
 three spatial directions. Dependence on spatial momenta and on the
 energy shows severe symmetry breaking effects. The vertices can be
 obtained from those of the Wilson case by
 \mbox{$r f_{\mu} \to i \lambda \gamma_{4}(1-\delta_{\mu 4}) f_{\mu}$}.
 This interaction does not change the temporal vertices at all!

 The self-energy is computed (for $\lambda=1$) in the same way as in the
 Bori\c ci-Creutz case:
 \begin{equation}
  \Sigma(p,m_{0}) = i \slashed{p} \Sigma_{1}(p) + m_{0} \Sigma_{2}(p)
  + c_{1}(g_{0}^{2}) i \cdot \frac{1}{2}(\slashed{p}+\slashed{p}^{\,\prime})  + c_{2}(g_{0}^{2}) i \frac{1}{a} \gamma_{4},
  \label{eq:KWselfenergy}
 \end{equation}
 with $\slashed{p}^{\,\prime} = \sum\limits_{\mu} p_{\mu} \gamma_{\mu}^{\,\prime}$,
 where $\gamma_{\mu}^{\,\prime}$ is defined in (\ref{eq:KWgammaprime}),
 $p_{4} \gamma_{4} \equiv \frac{1}{2}(\slashed{p}+\slashed{p}^{\,\prime})$ and
 \begin{eqnarray}
  & \Sigma_{1}(p) =
   & 1+\frac{g_{0}^{2}C_{F}}{16 \pi^{2}}\Big( \log a^{2}p^{2} + 9.2409
     +(1-\alpha) \big( -\log a^{2}p^{2}+4.792010 \big)\Big), \\
  & \Sigma_{2}(p) =
   & 1+\frac{g_{0}^{2}C_{F}}{16 \pi^{2}}\Big( 4\log a^{2}p^{2} -24.3688 
     +(1-\alpha) \big( -\log a^{2}p^{2}+5.792010 \big)\Big), \\
  & c_{1}(g_{0}^{2}) =
   & -0.1255 \cdot\frac{g_{0}^{2}C_{F}}{16 \pi^{2}}, \\
  & c_{2}(g_{0}^{2}) =
   & -29.5323 \cdot\frac{g_{0}^{2}C_{F}}{16 \pi^{2}}. \label{eq:KWc2}
 \end{eqnarray}
 The power-divergence in (\ref{eq:KWselfenergy}) concerns only the fourth
 direction. Otherwise, momentum renormalisation is handled analogously:
 $\check p_{4} = p_{4} -\frac{c_{2}(g_{0}^{2})}{a}$. We note in passing,
 that the definition of a rest frame is not problematic here. The magnitude
 of $c_{2}(g_{0}^{2})$ in (\ref{eq:KWc2}) is nearly the same as in the
 Bori\c ci-Creutz case (\ref{eq:BCc2}), but the sign has changed.

 The local bilinears (for $\lambda=1$) show similarities to the previous
 case. Exact chiral symmetry is maintained. The corrections to the proper
 vertices without wave-function renormalisation read
 \begin{eqnarray}
  & \Lambda_{S}(p) =
   & \!\frac{g_{0}^{2}C_{F}}{16 \pi^{2}}\Big(\! -4\log a^{2}\!p^{2} +24.36875
     +\!(1-\alpha) \!\big( \log a^{2}\!p^{2}-5.792010 \!\big)\!\Big), \\
  & \Lambda_{V}(p) =
   & \!\frac{g_{0}^{2}C_{F}}{16 \pi^{2}} \Big(\! -\log a^{2}\!p^{2} +10.44610 
     +\!(1-\alpha) \!\big( \log a^{2}\!p^{2}-4.792010 \!\big)\!\Big)\!, \\
  & \Lambda_{T}(p) =
   & \!\frac{g_{0}^{2}C_{F}}{16 \pi^{2}} \Big(\! 4.17551 
     +\!(1-\alpha) \!\big( \log a^{2}\!p^{2}-3.792010 \!\big)\!\Big).
 \end{eqnarray}
 Vector and axial currents suffer further operator mixing beyond
 the wave-function renormalisation
 \begin{equation}
  \overline{\psi} \,\gamma_{\mu} \psi \to
  \overline{\psi}^{R} \gamma_{\mu} \,\big(1+ Z_{\psi} + \Lambda_{V}(p) \big) \psi^{R}
  + c^{vtx}(g_{0}^{2})\overline{\psi}^{R} \,\gamma_{4} \psi^{R},
 \label{eq:KWlvc}
 \end{equation}
 with $c^{vtx}(g_{0}^{2})=-2.88914$ and $2\gamma_{4} = 2 \delta_{\mu4} \gamma_{\mu}
 = \gamma_{\mu}+\gamma_{\mu}^{\,\prime}$. This mixing can be taken into account
 more simply than in the Bori\c ci-Creutz case by choosing the spatial components'
 renormalisation factors different from the temporal one.

 Conserved currents are defined for Karsten-Wilczek fermions as well.
 The vector current reads
 \begin{eqnarray}
  V_{\mu}^{c}(x) = &
   \frac{1}{2} \Big(\overline{\psi}(x)(\gamma_{\mu}-i \gamma_{4}(1-\delta_{\mu 4}))
   U_{\mu}(x)\psi(x+a\, e_{\mu}) \nonumber\\&
   +\overline{\psi}(x+a\, e_{\mu})(\gamma_{\mu}+i \gamma_{4}(1-\delta_{\mu 4}))
   U_{\mu}^{\dagger}(x)\psi(x) \Big).
 \label{eq:KWcurrents}
 \end{eqnarray}
 The axial current $A_{\mu}^{c}$ is obtained by inserting $\gamma_{5}$
 behind the Dirac matrices. We have verified for $\lambda=1$ that their
 one-loop renormalisation factors are unity under analogous assumptions.

\section{Conclusions}

 We have demonstrated that the one-loop diagrams of Bori\c ci-Creutz fermions
 and Karsten-Wilczek fermions contain both similar structures.

 Mixings with marginal operators can be cast into forms which are clearly equivalent.
 Conserved point-split currents can be defined and their renormalisation factors are unity.

 Whereas in the former case all components of the four-momentum are subject
 to additive renormalisation, the latter case requires only renormalisation of the
 fourth component. The definition of a rest frame is problematic only in the
 former case. Thus we expect that the implementation of Karsten-Wilczek
 fermions poses less obstacles than Bori\c ci-Creutz fermions.

\end{document}